\documentclass{XrU2005}
\usepackage{graphicx}

\title{{\it XMM-Newton} observations of X-ray emission from Jupiter}
\author[1]{G. Branduardi-Raymont}
\affil[1]{Mullard Space Science Laboratory, University College London,
Holmbury St Mary, Dorking, Surrey RH5 6NT, UK}

\author[2]{A. Bhardwaj}
\affil[2]{Space Physics Laboratory, Vikram Sarabhai Space Centre, 
Trivandrum 695022, India}

\author[3]{R. F. Elsner}
\affil[3]{NASA Marshall Space Flight Center, NSSTC/XD12, 320 Sparkman Drive, 
Huntsville, AL 35805, USA}

\author[4]{G. R. Gladstone}
\affil[4]{Southwest Research Institute, P. O. Drawer 28510, San Antonio, 
Texas 78228, USA}

\author[1]{G. Ramsay}

\author[5]{P. Rodriguez}
\affil[5]{XMM-Newton SOC, Apartado 50727, Villafranca, 28080 Madrid, Spain}

\author[1]{R. Soria}

\author[6]{J. H. Waite, Jr}
\affil[6]{University of Michigan, Space Research Building, 2455 Hayward, Ann 
Arbor, Michigan 48109, USA}

\author[7]{T. E. Cravens}
\affil[7]{Department of Physics and Astronomy, University of Kansas, Lawrence, 
KS 66045, USA}

\begin{document}


\maketitle

\begin{abstract}
We present the results of two {\it XMM-Newton} observations of Jupiter
carried out in 2003 for 100 and 250 ks (or 3 and 7 planet rotations)
respectively. X-ray images from the EPIC CCD cameras show prominent emission
from the auroral regions in the 0.2$-$2.0 keV band: the spectra
are well modelled by a combination of emission lines, including most
prominently those of highly ionised oxygen (OVII and OVIII). In addition, and
for the first time, {\it XMM-Newton} reveals the presence in both aurorae 
of a higher energy component (3$-$7 keV) which is well described by an 
electron bremsstrahlung spectrum. This component is found to be variable
in flux and spectral shape during the Nov. 2003 observation, which 
corresponded to an extended period of intense solar activity. Emission from 
the equatorial regions of the Jupiter's disk is also observed, with a spectrum 
consistent with that of solar X-rays scattered in the planet's upper 
atmosphere. Jupiter's X-rays are spectrally resolved with the RGS which
clearly separates the prominent OVII contribution of the aurorae from the 
OVIII, FeXVII and MgXI lines, originating in the low-latitude disk regions 
of the planet.
\end{abstract}

\section{Introduction}

Jupiter was first detected as an X-ray source with the {\it Einstein} 
observatory \citep{Met:83}. By analogy with the Earth's aurorae, the 
emission was expected to be produced via bremsstrahlung by energetic
electrons precipitating from the magnetosphere. However, the observed 
spectrum is softer (0.2$-$3 keV) and the observed fluxes larger than predicted
from this mechanism. Model calculations by \citet{Sing:92} 
confirmed that the expected bremsstrahlung flux is lower by 1 to 2
orders of magnitude compared with the observed $<$2 keV X-ray flux.
The alternative process is K shell line emission from 
ions, mostly of oxygen, which charge exchange, are left in an excited
state and then decay back to the ground state (see Bhardwaj and Gladstone, 
2000, for a review of early work on planetary auroral emissions). The ions 
were thought to originate in Jupiter's inner magnetosphere, where an
abundance of sulphur and oxygen, associated with Io and its plasma torus,
is expected \citep{Met:83}. 

The first {\it ROSAT} soft X-ray (0.1$-$2.0 keV) observations produced a 
spectrum much more consistent with recombination line emission than with 
bremsstrahlung (Waite et al., 1994; Cravens et al., 1995). Subsequent
{\it ROSAT} observations also revealed low-latitude `disk' emission from 
Jupiter \citep{Waite:97}, and this too was attributed to charge exchange. 
However, the X-rays were brightest at the planet's limb corresponding
to the position of the subsolar point relative to the sub-Earth point,
suggesting that a solar-driven mechanism may be at work 
\citep{gla:98}. Scattering of solar X-rays, both elastic (by atmospheric
neutrals) and fluorescent (of carbon K-shell X-rays off methane molecules below
the Jovian homopause), was put forward as a way to explain the disk emission
\citep{mau:00}.

With the advent of the {\it Chandra} observatory we gained the clearest view
yet of Jupiter's X-ray emission, but more questions arose as well: HRC-I
observations in Dec. 2000 and Feb. 2003 clearly resolve two bright,
high-latitude sources associated with the aurorae, as well as low-latitude
emission from the planet's disk (Gladstone et al., 2002; Elsner et al., 2005).
However,
the Northern X-ray hot spot is found to be magnetically mapped to distances
in excess of 30 Jovian radii, rather than to the inner
magnetosphere and the Io plasma torus. Since in
the outer magnetosphere ion fluxes are insufficient to explain the observed
X-ray emission, another ion source (solar wind?) and/or acceleration
mechanism are required. Strong 45 min quasi-periodic X-ray oscillations were
also discovered using {\it Chandra} data in the North auroral spot in 
Dec. 2000,
without any correlated periodicity being observed in {\it Cassini} upstream
solar wind data, or in {\it Galileo} and {\it Cassini} energetic particle and
plasma wave measurements \citep{gla:02}. 
{\it Chandra} ACIS-S observations \citep{els:05} show that the auroral X-ray 
spectrum is made up of oxygen line emission consistent with mostly fully 
stripped ions. 
Line emission at lower energies could be from sulphur and/or carbon. The high 
charge states and the observed fluxes imply that the ions must have undergone 
acceleration, 
independently from their origin, magnetospheric or solar wind. Rather than 
periodic oscillations, chaotic variability of the auroral X-ray emission was 
observed, with power peaks in the 20$-$70 min range. 
A promising mechanism which could explain this change in character of the 
variability, from organised to chaotic, is pulsed reconnection at the 
day-side magnetopause, as suggested by \citet{bun:04}. 

\section{{\it XMM-Newton} observations}

\begin{figure}
\centering
\includegraphics[height=0.8\linewidth]{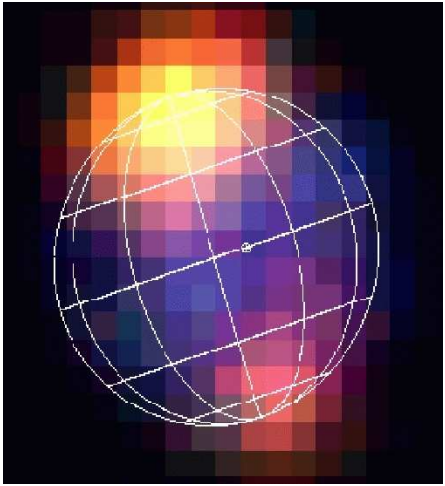}
\vspace{-0.2cm}
\caption{
Smoothed {\it XMM-Newton} EPIC image of Jupiter (2.9''\,pixels),
Apr. 2003. North is to the top, and East to the left. 
Colour code: Red: 0.2--0.5 keV; Green: 0.5--0.7 keV; Blue: 0.7--2.0
keV. The equatorial emission is clearly harder than that from the auroral
regions. A graticule showing Jupiter orientation with 30$^{\rm o}$ intervals 
in latitude and longitude is overlaid. The circular mark indicates the 
sub-solar point; the sub-Earth point is at the centre of the graticule.
\label{fig1:single}
}
\end{figure}

{\it XMM-Newton} observed Jupiter twice in 2003: between Apr. 28, 16:00
and Apr. 29: 22:00 UT (for a total observing time of 110 ks; see
Fig.~\ref{fig1:single}, from 
Branduardi-Raymont et al., 2004, BR1 hereafter), and between Nov. 25, 23:00
and Nov. 29, 12:00 UT (245 ks, split over two spacecraft revolutions, no.s
0726 and 0727; Branduardi-Raymont et al., 2006a,b). 

\subsection{Temporal behaviour}

\begin{figure}
\centering
\hspace{-0.5cm}
\includegraphics[angle=90,width=1.05\linewidth]{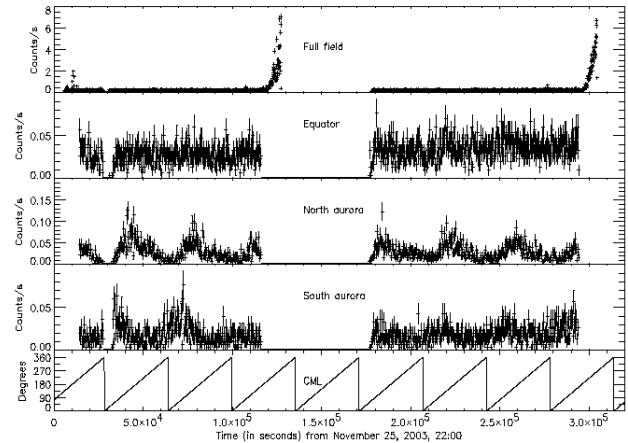}
\caption{Jupiter lightcurves from the Nov. 2003 {\it XMM-Newton} observation. 
Middle three panels: Low-latitude disk and auroral emissions (0.2--2.0 keV, 
300 s bins). Top panel: Lightcurve at energies $>$10 keV, 
showing periods of high background (excluded from the analysis) at the end 
of the two spacecraft orbits. Bottom panel: System III Central
Meridian Longitude (CML). 
\label{fig2:single}
}
\end{figure}

Lightcurves from the Nov. 2003 observation, shown in Fig.~\ref{fig2:single}, 
resemble very closely those obtained the previous April (BR1). 
The planet 10 hr rotation period is clearly seen in the data
of the North and South auroral spots, but not in the equatorial region.
The bottom panel in Fig.~\ref{fig2:single} shows the System III Central 
Meridian Longitude (CML). The North spot is brightest around CML = 
180$^{\rm o}$, similar to the Dec. 2000 {\it Chandra} and the Apr. 2003 
{\it XMM-Newton} results. A 40\% increase in the equatorial flux 
between the first and the second spacecraft revolution is noticeable in 
Fig.~\ref{fig2:single}, and is found to be correlated with a similar
increase in solar X-ray flux (see Bhardwaj et al., 2005, for 
a detailed study of the temporal behaviour of the low-latitude disk emission,
which appears to be controlled by the Sun). A search for periodic
behaviour on short time-scales in the auroral X-rays (i.e. the {\it Chandra}
45 min oscillations) leads to a null result (as for the Apr. 2003 
{\it XMM-Newton} data). This supports the view that over time the character of 
the variability in the auroral X-ray emissions can change from well 
organised to chaotic.

\subsection{EPIC spectral images}

The {\it XMM-Newton} observation of Jupiter in Apr. 2003 
gave the first clear indication that the Jovian auroral and disk X-ray 
emissions have different spectra. Fig.~\ref{fig1:single} (BR1) 
is the planet's image colour-coded depending on X-ray energy: the 
equatorial disk emission is clearly harder 
that that of the aurorae. The auroral spectra can be modelled 
with a superposition 
of emission lines, including most prominently those of highly ionised 
oxygen (OVII and OVIII). Instead, Jupiter's low-latitude X-ray 
emission displays a spectrum consistent with that of solar 
X-rays scattered in the planet's upper atmosphere (BR1). 
These results are strengthened by the Nov. 2003 observation. 

\begin{figure*}
\centering
\includegraphics[angle=-90,width=0.8\linewidth]{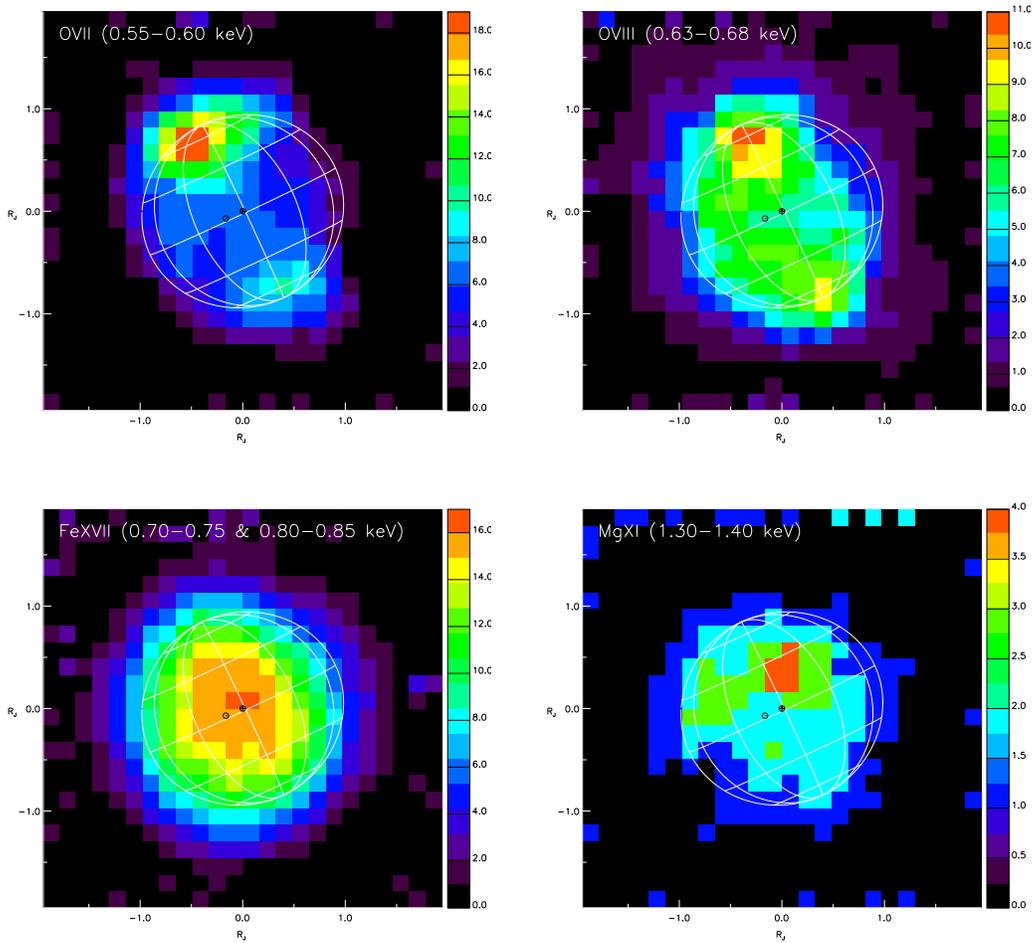}
\caption{Smoothed {\it XMM-Newton} EPIC images of Jupiter in narrow
spectral bands. From top left, clockwise: OVII, OVIII, MgXI, FeXVII. 
The colour scale bar is in units of EPIC counts.
\label{fig3:double}
}
\end{figure*}

Figs.~\ref{fig3:double} shows the combined EPIC-pn \citep{stru:01} and -MOS
\citep{tur:01} CCD images in narrow spectral 
bands corresponding to the OVII, OVIII, FeXVII and MgXI lines detected in 
Jupiter's spectra: OVII emission is 
concentrated mostly in the North and (more weakly) the South auroral spots, 
OVIII extends to lower latitudes, while FeXVII and MgXI display a rather 
uniform distribution over the planet's disk, consistent with an origin 
from scattered solar X-rays.

\begin{figure*}
\centering
\includegraphics[angle=-90,width=0.8\linewidth]{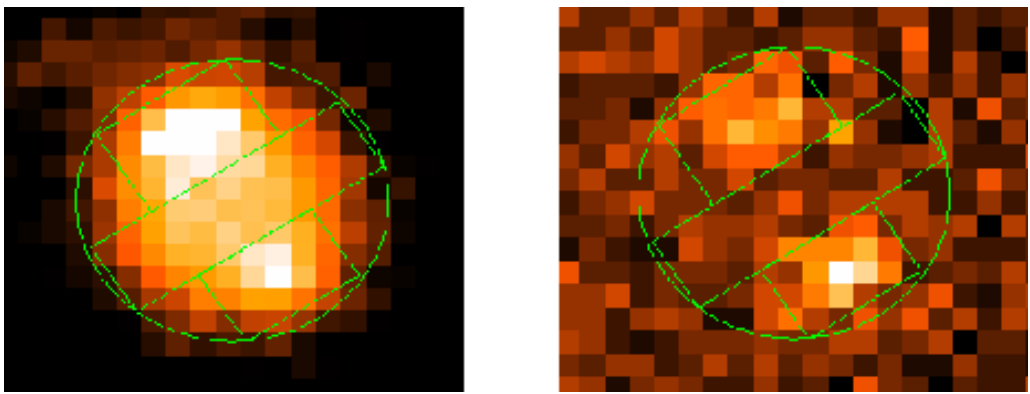}
\caption{Jupiter's images from the combined {\it XMM-Newton} EPIC
cameras data ($\sim$1.4\arcmin\ side; left: 0.2$-$2 keV band; right: 3$-$10 
keV). Superposed are the regions used to extract auroral and low-latitude 
disk lightcurves and spectra.
\label{fig4:double}
}
\end{figure*}

Although most of the X-ray emission of Jupiter is
confined to the 0.2$-$2 keV band, a search at higher energies has
produced very interesting results. Fig.~\ref{fig4:double} (right) is an image 
of Jupiter
in the 3$-$10 keV band, which shows the presence of higher energy emission from
the auroral spots, but not so from the planet's disk. 

\subsection{EPIC spectra}

\begin{figure}
\centering
\includegraphics[angle=-90,width=1.0\linewidth]{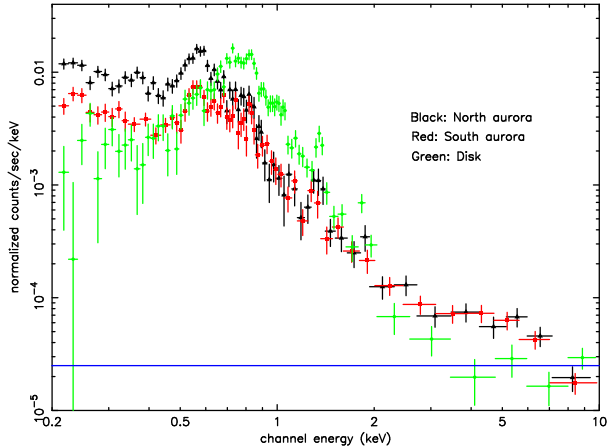}
\caption{Combined EPIC spectra of the North (black) and South (red)
aurorae, and of the low-latitude disk (green) spectrum. 
Differences in spectral shape 
between auroral and disk spectra are clear. The presence of a high energy
component in the spectra of the aurorae is very evident, with a substantial
excess relative to the disk emission extending to 7 keV. The horizontal blue 
line shows the estimated level of the EPIC particle background.
\label{fig6:single}
}
\end{figure}

EPIC CCD spectra of Jupiter's auroral zones and low-latitude disk
emission were extracted using the regions outlined in Fig.~\ref{fig4:double};
the spectral `mixing' (due to the {\it
XMM-Newton} Point Spread Function) was corrected for by subtracting
appropriate fractions of disk and auroral emissions from the aurorae and 
the disk spectra respectively.
Fig.~\ref{fig6:single} compares the resulting spectra of the North and South
auroral spots and the disk for the Nov. 2003 observation. 

\begin{figure}
\centering
\includegraphics[angle=-90,width=1.0\linewidth]{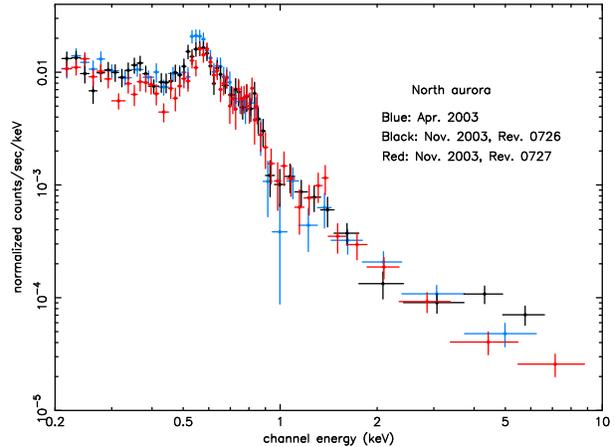}
\caption{Combined EPIC spectra of the North aurora for the two 
separate {\it XMM-Newton} revolutions, 0726 (black) and 0727 (red), in
Nov. 2003, and for the Apr. 2003 observation (blue).
\label{fig7:single}
}
\end{figure}

As first pointed out by BR1, there are clear
differences in the shape of the spectra, with the auroral emission peaking
at lower energy (0.5$-$0.6 keV) than the disk (0.7$-$0.8 keV). Emission
features in the range 1$-$2 keV are visible in all the spectra, but are
stronger in the disk \citep{bra:06b}. The presence
of a high energy component from the aurorae is confirmed, while
this is missing in the disk emission. Variability in the auroral spectra is 
also observed (Fig.~\ref{fig7:single}): the high energy 
part of the auroral spectra varies between the two Nov. 2003 {\it XMM-Newton}
revolutions, and changes are also observed at low energies.

\subsection{EPIC spectral fits}

\begin{figure}
\centering
\includegraphics[angle=-90,width=1.0\linewidth]{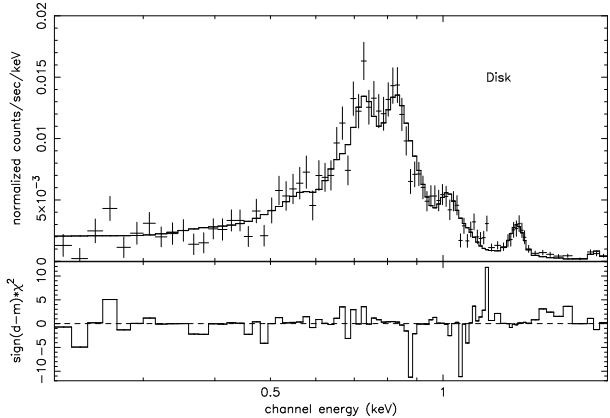}
\caption{EPIC spectrum of Jupiter's disk and {\tt mekal} 
(collisional plasma model) best fit (see text for details).
\label{fig8:single}
}
\end{figure}

A collisional plasma model ({\tt mekal} in XSPEC) with temperature
kT = 0.46 $\pm$ 0.03 keV is a good representation of the low-latitude 
disk spectrum (Fig.~\ref{fig8:single}), after including additional 
MgXI and SiXIII emission (at 1.35 and 1.86 keV respectively, likely
consequences of enhanced solar activity) and a small contribution of OVII
(0.57 keV) and OVIII (0.65 keV, both residual auroral contamination).

\begin{figure}
\centering
\includegraphics[angle=-90,width=1.0\linewidth]{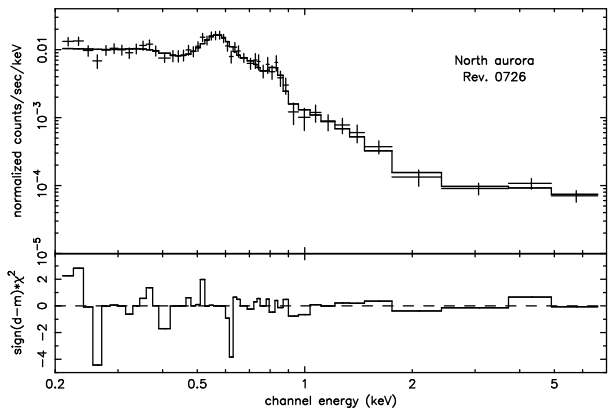}
\caption{{\it XMM-Newton} EPIC spectrum of Jupiter's North aurora
from the Nov. 2003, rev. 0726 observation, fitted with thermal bremsstrahlung 
and power law continua, plus four lines (see text for details).
\label{fig9:single}
}
\end{figure}

The auroral spectra are well fitted by a model comprising two thermal 
bremsstrahlung
continua and four gaussian emission lines: these are found at 0.32 keV (C
and/or S), 0.57 (OVII), 0.69 (OVII and FeXVII) and 0.83 keV (Fe XVII) for 
the rev. 0726 spectra; in rev. 0727 and in Apr. 2003 the lowest energy 
line is not present but one is needed at 1.35 keV (MgXI, probably 
residual contamination from scattered solar X-rays). 
The bremsstrahlung continua reflect the presence of two distinct spectral 
components dominating at the low and high energy end respectively. The 
temperature of the low energy component is fairly
stable, ranging between 0.1 and 0.3 keV and being practically the same for 
both aurorae. For the higher energy component, the rev. 0726 spectra 
require a much higher bremsstrahlung temperature than those from the two 
other epochs. At the same time the addition of an emission line is needed
in order to explain a peak at 0.3$-$0.4 keV. 
In actual fact, the spectral shape at the higher energies is better matched
by a very flat power law (photon index $\sim$0.2) than a hot thermal
bremsstrahlung. The spectrum and best fit for the North aurora from 
the Nov. 2003, rev. 0726 observation are shown in Fig.~\ref{fig9:single}.

\begin{figure}
\centering
\includegraphics[angle=-90,width=0.8\linewidth]{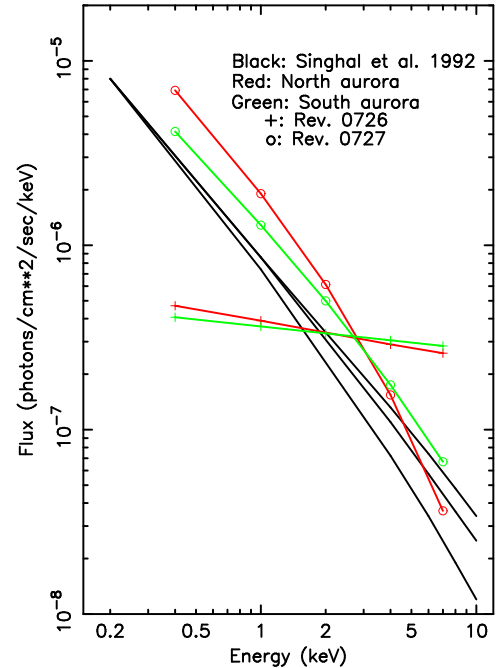}
\caption{High energy model components fitted to the Nov. 2003 auroral 
data, compared with Singhal et al. (1992) bremsstrahlung X-ray 
flux predictions for three characteristic electron energies (10, 30 and 
100 keV, from bottom to top curve).
\label{fig10:single}
}
\end{figure}

Fig.~\ref{fig10:single} displays the high energy continuum model components 
fitted to the Nov. 2003 auroral data (flat power law for rev. 0726 and steeper
bremsstrahlung for rev. 0727) and compares them with the predictions of
Singhal et al. (1992) for bremsstrahlung emissions by electrons
of energies between 10 and 100 keV. The bremsstrahlung fit of rev. 0727
lies remarkably close to the predicted spectrum for both the North and South
aurorae. The models for rev. 0726, 
however, suggest a very different electron distribution for both aurorae. 

\subsection{RGS spectrum}

\begin{figure}
\centering
\vspace{-0.6cm}
\includegraphics[angle=-90,width=1.0\linewidth]{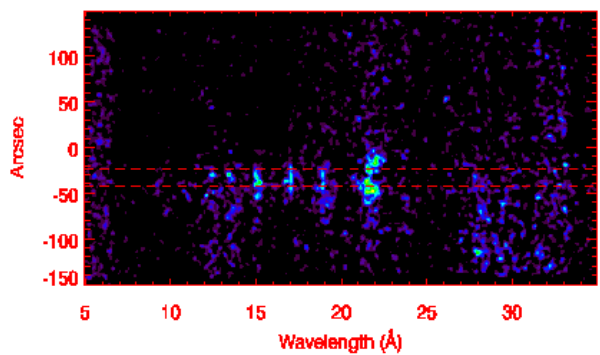}
\caption{RGS spectrum of Jupiter from the combined RGS1 and 2 datasets
of both {\it XMM-Newton} revolutions in Nov. 2003; the
emission spatial extent in the cross dispersion direction is visible
along the vertical axis, while the X-ray wavelength
is plotted along the horizontal axis. The two dashed horizontal lines mark
the location of Jupiter's aurorae (the planet's N$-$S axis was essentially 
perpendicular to the RGS dispersion direction).
\label{fig11:single}
}
\end{figure}

\begin{figure}
\centering
\includegraphics[angle=-90,width=1.0\linewidth]{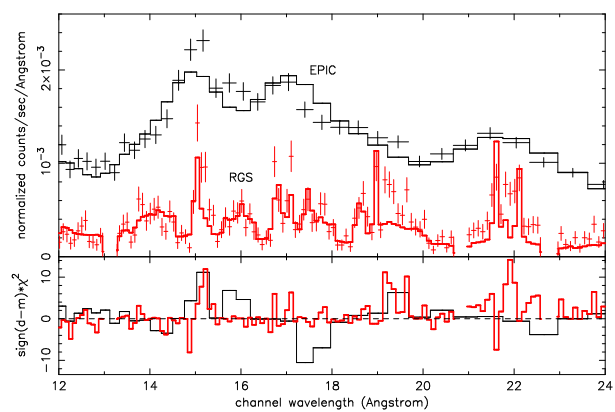}
\caption{EPIC and RGS spectra of Jupiter (full planet)
and simultaneous best fit with a combination of {\tt mekal} models.
\label{fig12:single}
}
\end{figure}

Fig.~\ref{fig11:single} shows the RGS spectrum of Jupiter obtained by coadding
the RGS1 and 2 data (first order only) from both {\it XMM-Newton} 
revolutions in Nov. 2003: the image
is colour-coded according to the detected flux, and displays the spatial 
distribution of the emission in the cross dispersion direction versus 
X-ray wavelength.
The RGS clearly separates the emission lines of OVII (21.6$-$22.1 \AA, or
0.56$-$0.57 keV), OVIII (19.0 \AA, or 0.65 keV) and FeXVII (15.0 and
$\sim$17.0 \AA, or $\sim$0.73 and 0.83 keV). 
Interestingly, the RGS spectrum also shows evidence for the different spatial 
extension of the line emitting regions, in agreement with the EPIC spectral 
mapping of Fig.~\ref{fig3:double}: OVII photons are well separated into the 
two aurorae, while the other lines are filling in the low latitude/cross 
dispersion range. 
The higher resolution RGS spectrum, which includes X-ray light from the whole 
planet, agrees well, in flux and profile, with the EPIC one integrated over 
the full disk of Jupiter (Fig.~\ref{fig12:single}). 

\section{Discussion and Conclusions}

{\it XMM-Newton} observations of Jupiter on two epochs in Apr. and Nov. 2003
convincingly demonstrate that auroral and low-latitude disk X-ray emissions
are different in spectral shape and origin. The Jovian auroral soft X-rays
($<$ 2 keV) are most likely due to charge exchange by energetic ions from 
the outer magnetosphere, or solar wind, or both. For the first time a 
higher energy component in the auroral spectra has been identified, and 
has been found to be variable over timescales of days: its spectral shape is 
consistent with that predicted from bremsstrahlung of energetic electrons
precipitating from the magnetosphere. The variability observed in its flux and 
spectrum is likely to be linked to changes in the energy distribution of the 
electrons producing it and may be related to the particular period of intense 
solar activity reported in Oct. - Nov. 2003 by a number of spacecraft 
mesurements.

\section*{Acknowledgments}

This work is based on observations obtained with {\it XMM-Newton}, an ESA
science mission with instruments and contributions directly funded by ESA
Member States and the USA (NASA). The MSSL authors acknowledge financial 
support from PPARC.

\end{document}